\documentclass{article}
\usepackage{amssymb}
\usepackage{makeidx}
\usepackage{amsfonts}
\usepackage{amsmath}
\usepackage{color}
\usepackage{graphicx}

\newtheorem{corollary}{Corollary}

\newtheorem{lemma}{Lemma}

\newtheorem{proposition}{Proposition}

\newtheorem{assumption}{Assumption}

\newenvironment{proof}[1][Proof]{\noindent\textbf{#1.} }{\ \rule{0.5em}{0.5em}}

\begin{document}

\title{ Unleashing the Transformative Power of Deliberation With Contextual Citizens}
\vspace{0.3cm}

\author{Ariane Lambert-Mogiliansky\thanks{%
Paris School of Economics, 48 Boulevard Jourdan, Paris 75014.
(alambert@pse.ens.fr)} \thanks{%
I want to thank Fran\c{c}ois Dubois et Zeno Toffano for very useful discussions and suggestions in
the early stages of the project. } and
 Ir\'{e}n\'{e}e Fr\'{e}rot \thanks{%
Laboratoire Kastler, Brossel, Sorbonne Universit\'{e}, CNRS, ENS-PSL
Research University, Coll\`{e}ge de France, 4 Place Jussieu, 75005 Paris,
France. (irenee.frerot@lkb.upmc.fr)}}
\maketitle

\begin{abstract}
In this paper, we investigate deliberation procedures that invite citizens with contextual opinions to explore alternative thinking frames. Contextuality is captured in a simple quantum cognitive model. We show how disagreeing citizens endowed with contextual opinions, can reach consensus in a binary collective decision problem with no improvement in their information. A necessary condition is that they are willing to (mentally) experience their fellow citizens' way of thinking.  The diversity of thinking frames is what makes it possible to overcome initial disagreement. Consensus does not emerge spontaneously from deliberations: it requires facilitation. \\

\end{abstract}

\section{Introduction}

Recently, representative democracy has been questioned and is widely
perceived as being in crisis in most developed countries. At the same time,
more participative forms of democracy are gaining interest \cite{Science19}. As Niemeyer \textit{et al.} write: ``Deliberative democracy is now arguably the main theme in
both democratic theory and the practice of democratic innovations" \cite{niemeyer24}. Theories of collective decision-making are traditionally
partitioned into two major fields: those dealing with issues related to the
aggregation of diverse preferences (Social choice) and those dealing with
citizen's active participation aiming at fostering reciprocal understanding
and compromise and move toward consensus (see \cite{list2018} for a review)%
\footnote{%
Deliberative institutional experimentation is flourishing throughout the
world (a catalog is available at https://participedia.net/).}.

 For participatory democrats
from John Stuart\ Mills to Carole Pateman, the goal of politics is the
transformation and education of participants, so that politics is an end in
itself \cite{Elster05}. Data from Deliberative Polls support the
hypothesis that people do change their opinion, and this happens not only
under the impact of better information \cite{Listetal13,Farrar10}.
J. Dryzek writes that a \textquotedblleft defining feature of deliberative
democracy is that individuals participating in democratic processes are
amenable to changing their minds and their preferences as a result of the
reflection induced by deliberation\textquotedblright \cite{dryzek02}.

Deliberative democratic theory relies on the principle that
\textquotedblleft outcomes are democratically legitimate if and only if they
could be the object of a free and reasoned agreement among
equals\textquotedblright \cite{Cohen97}. But how does the process of presenting arguments leads to agreement among equals? Some
scholars have argued that reasoned public deliberation lends legitimacy
because the proposals that are sustained and survive are simply better in terms of their overall quality. This presumes that there exist some procedure independent
criteria of rightness or correctness. Many epistemic democrats hold this view \cite{estlund97,lande13}. Others have proposed that
the very procedure of reasoned public
deliberation embodies or manifests core values of basic human morality and
political justice, and it forces participants to be attentive toward the
common good \cite{Christ97,Cohen97,Knight97,awls97}. Finally, a number of scholars have argued that reasoned public
deliberation complements (or even nullify the need for)
voting mechanisms by \textquotedblleft
inducing a shared understanding regarding the dimensions of
conflict\textquotedblright \cite{Knightjohn94} which prevents the majority rule from generating majority cycles \cite{dryzeklist03,Listetal13}). Bohman emphasizes both the
transformative and epistemic benefits of confronting a diversity of
perspectives in deliberations \cite{Bohman06}. Our approach is, in its
spirit, close to that of Bohman's. We view the process of deliberation as a
procedure that invites citizens to explore alternative perspectives.

The central hypotheses of this paper is: i. To be able to consider an
issue, people have to build a representation of that issue. Building a
representation requires selecting a perspective, or equivalently a thinking frame, namely a model; and ii. There exist perspectives that people cannot consider
simultaneously, that are incompatible in the mind. This has the crucial
implication that no single perspective can aggregate all relevant
information: opinions are contextual. In a close spirit Niemeyer \textit{et al.}
write: ``Deliberative reasoning as we characterize it, recognizes the
possibility of identifying the set of relevant considerations, while falling
short by failing actively to take all of them into account to capture the
complete picture" (\cite{niemeyer24} p. 347). To focus on the evolution of
opinions due to their contextuality, we consider deliberation exclusively as
a process of confronting alternative models with no improvement in
information. Kinder writes: ``frames supply no new information.
Rather, by offering a particular perspective, frames \textit{organize - or
better reorganize - }information that citizens already have in mind"\cite%
{kinder03}. Frames suggest how politics should be thought about, encouraging
citizens to think in particular ways". 
To address the
contextuality of opinions, we turn to a widely recognized formal
approach that features the co-existence of alternative representations of
one and the same object: the Hilbert space model of Quantum Mechanics (QM). Under the last decades, quantum-like
models of contextuality have been developed in Social Sciences to explain a
variety of behavioral anomalies (for overviews, see refs.~\cite%
{bubu12}, \cite{Yearbu16}, \cite{Jerry20}). We briefly
introduce the quantum cognition approach in Section \ref{sec_contextuality} and then more formally in Section \ref{sec_model}. We emphasize that no prior knowledge of QM or Hilbert space is needed
to read the present paper.

We consider a setting where there is a Yes or No decision to be made%
\footnote{%
This implies that we are not addressing
the impossibility theorems of collective choice \cite{Arrow1950} }.
The objective of deliberation is to achieve democratic legitimacy in the
following sense. First, the procedure should give a fair chance to
everyone's opinion to affect the decision. Second, full consensus is the
overarching goal which translates in maximizing the support for the final
decision. In this context, the paper addresses two central questions: 1. How
can (fact-free i.e., without additional information) deliberation affect
citizens' voting behavior? 2. How should we structure deliberation to maximize
the probability for consensus? A central assumption is that citizens are willing to explore
alternative thinking frames before deciding how to vote. These alternatives
frames can be provided by the citizens themselves and/or experts. The
procedure is managed by a benevolent facilitator. We find that in the two-person case (Section \ref{sec_2citizens}), when
starting from opposite voting intentions, fact-free deliberation 
always achieves some extent of consensus provided that the two citizens do not share the same thinking frame at the outset: diversity of viewpoints is beneficial. The largest chance of consensus is obtained when the citizens' perspective are maximally uncorrelated\footnote{Perspective A and B
are said to be maximally uncorrelated whenever a citizen's opinion in
perspective A has no predictive power for her opinion in perspective
B.}. In the case where the perspectives are two-dimensional
consensus can be reached with probability $3/4$.
 In Section \ref{sec_population}, we consider a
population of citizens and find that to maximize the speed of convergence to consensus, the
facilitator should introduce some asymmetries between citizens, and in some cases overturning the initial majority may turn out to be optimal. In the Section \ref{sec_discussion} 
we address some epistemic considerations and the performance of our model with
respect to empirical studies. 

The main takeaways from our account of
thinking frames in the analysis of fact-free deliberations are: i. The diversity of
thinking frames among citizens, not only is no obstacle, but is a necessary
condition for deliberation to deliver consensus; ii. Deliberation can exhibit a
transformative power that hinges on a willingness of participants to explore alternative thinking frames; iii. Well-designed procedures
monitored by a facilitator have a significant potential to help people reach consensus.

This paper contributes to the literature on the value of pre-voting
deliberation by providing a formalisation of opinion formation appealing to
the intrinsic contextuality of opinions. Most formal approaches to deliberation belong to the epistemic tradition
which postulates a single (common) correct decision. Among the most recent
ones Dietrich and Spiekermann introduce some behavioral dimensions (``sharing
and absorbing") to the information theoretic approaches (see e.g., \cite%
{Dietrich24}). The other formal strand of literature is game-theoretic. It
emphasizes incentives to share or withhold information (see e.g., \cite{piva21}%
). The quantum (contextuality) revolution has instead recasted the issue of objective
truth and knowledge, as witnessed by the wealth of the epistemological
literature over the last century (see e.g., \cite{despa12}).\ Our approach
based on the most standard formalisation of intrinsic contextuality is
closely related to Bohman's experiential perspective approach in \cite%
{Bohman06}. He emphasizes the transformative and epistemic benefits of
confronting the diversity of perspectives in deliberations. We provide a formal
description for this transformation and derive optimal procedures managed by a
facilitator. Because we preclude
improvement in information, our approach is complementary to classical epistemic
approaches. But since intrinsic contextuality precludes the uniqueness of truth, it bring us close to procedural approaches to deliberation
\cite{beauvais16} which recognize a value to deliberation
in terms of realizing basic values of
democracy. Our results also contribute with new results with respect to the 
tension between efficiency and legitimacy within the process of deliberations itself \cite{dryzek01}.

\section{Contextuality in social sciences}
\label{sec_contextuality}

\begin{figure}
    \centering
    \includegraphics[width=0.5\linewidth]{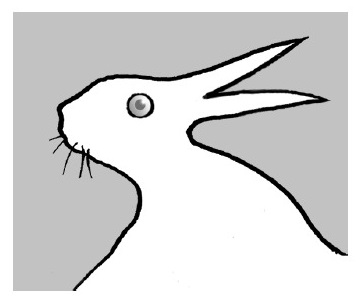}
    \caption{What do you see?}
    \label{fig_lapin}
\end{figure}

It is a common place that human beings are not capable of holding very
complex pictures in mind. We consider reality focusing on one perspective (or thinking frame) at
a time and show difficulties in combining perspectives in a stable way. This
inability to seize reality in its full richness suggests that the process of
developing an understanding of the world may not look like a puzzle that is
assembled progressively. Instead, the human mind may exhibit structural
``limitations" in terms of the incompatibility of perspectives. This is reminiscent of the description of physical objects in quantum mechanics. The psychological process of learning is then better described as an explorative and transformative journey through different landscapes. Ambiguous pictures such as Fig.~\ref{fig_lapin} provides a suggestive illustration of the kind of incompatibility we have in mind.
You may see a duck, or a rabbit. You may oscillate between the two: both are correct, but you cannot see both simultaneously.

At first sight, it may appear somewhat artificial to turn to quantum mechanics (QM)
when investigating human behavioral phenomena. However, the founders of QM,
including Bohr and Heisenberg recognized an essential similarity between the
two fields\footnote{%
In particular, Bohr was influenced by the psychology and philosophy of
knowledge of Harald H\"{o}ffding (see ref. \cite{Bohr91} and the Introduction in
ref. \cite{bitbol09} for an insightful discussion).}. In both fields the object of
investigation cannot always be separated from the process of
investigation. QM and in particular its mathematical formalism was developed
to respond to a general epistemological challenge: how can one study an
object that is being modified in an uncontrollable way in the process of measurement of its properties? QM is a
general paradigm for intrinsic contextuality (i.e., non-separability between
the object and the operation of investigation). It is therefore fully legitimate to explore the value of the mathematical
formalism of QM in the study of human behavioral phenomena - without
reference to Physics.

The quantum paradigm has been proposed in decision theory and psychology as a \textit{quantum cognitive} approach to
describe preferences, beliefs, attitudes, judgements and opinions (see \cite%
{bubu12}, \cite{Yearbu16}, \cite{dan18c},\ \cite{Jerry20}, \cite{Hakhre13}, \cite{ALMZZ09} among others). Quantum Cognition has allowed for providing a unified framework that accommodates a large
variety of so-called behavioral anomalies: preference reversal,
disjunction effect, cognitive dissonance, framing effects, order effects, among others (see e.g. \cite{ALMZZ09}, and for empirical applications see \cite%
{emp09,calm21}).

In this paper, our starting point is that the representation of the world is a mental
object which may exhibit non-classical features, and we
derive some implications for the dynamics of opinion
formation in the process of deliberation. The most important element that we
borrow from QM is the notion of (Bohr-)complementarity, applied here to mental
perspectives.\footnote{%
It parallels the complementarity of properties for sub-atomic particles
e.g., the spin along different directions}. In line with Quantum Cognition, we propose that Bohr
complementarity of perspectives captures the cognitive limitations that are
responsible for our difficulties to synthesize information along different
perspectives into a single stable picture. Just as in QM, the system (here our
mental picture) makes discrete jumps when attempting to find a determinate
value along distinct incompatible perspectives (see again Fig.~\ref{fig_lapin}).

Recently, theoretical and experimental applications to persuasion have shown
how fruitful this approach could be \cite{dan18a,dan18b,calm21}. This paper is in continuation with those works. A citizen's
opinion is formalized as a quantum object characterized by its state vector.
Alternative thinking frames are modelled as alternative bases, representing incompatible perspectives, on the (mental representation of the) decision issue. Deliberation amounts to a sequence of
measurements (probing arguments). Measurements moves the opinion state in a
non-deterministic way that reflects the correlations between the perspectives (the
thinking frames). With this formal description of the process of
deliberation, we investigate procedures that satisfy some requirements put forward in the literature on democratic
deliberations. 

We emphasize that the quantum cognition approach does not assume a quantum
physical nature of the determinants of our opinions. Neither do we dwell
into the psychology or neurology of the transformation of belief/opinion and
preferences\footnote{%
The approach is an abstract way of capturing the fact that experiencing
another perspective may have neurological, emotional and other impacts with
consequences for opinion.}. A presumption is that the correlations between
perspectives which structure the mind exhibit some extent of regularity
across individuals. The quantification of such correlations remains an
empirical question open to future work.

\section{Model}
\label{sec_model}
\subsection{Basic structure}

We formulate our model of deliberation in terms of a mediated
communication game. We consider a set $\Omega $ of deliberator-citizens, one
facilitator and a pool of experts. We are interested in
deliberations aimed at influencing citizens vote over uncertain options
which we model as quantum lotteries, following ref.~\cite{dan18c}. The formal model shares significant features with the quantum persuasion
model of refs.~\cite{dan18a,dan18b}.

\begin{assumption}
\label{assum_0}
Citizens taking part in deliberation are willing to explore alternative thinking frames. They follow the recommendations of the facilitator.
\end{assumption}

Citizens participating in deliberations are assumed to be open-minded, i.e. they are willing to engage in real mental experiences. This captures a leitmotiv in the literature on deliberations, which recognizes the value of reciprocal respect and goodwill. The facilitator's recommendations consist exclusively in inviting a citizen or expert to present an argument, or in inviting some citizens to probe a perspective
(see below for a definition). In this paper, we are not addressing possible
incentive issues related to these operations. Instead, we assume benevolence from the side of the facilitator.\footnote{How to safeguard against potential manipulations  is a subject of great importance that is not addressed} here. \par \medskip
Throughout the paper, we shall illustrate the concepts with the following example:\par  
\textit{A community needs to decide whether or not to
introduce an Individual Carbon Budget (ICB). To
be able to relate to the issue, citizens build a mental representation using a thinking frame. Various aspects are of relevance to citizens, environmental
efficacy, impact on individual liberties, legal feasibility etc... each corresponding to a specific thinking frame.} \par \medskip

\subsection{Opinions and perspectives}

A citizen's opinion is a mental object that we model as a quantum system.\par 
The description of a quantum system starts with the definition of a Hilbert
space $H$ (over the field $\mathbb{R}$ of real numbers or the field $\ $of
complex numbers $\mathbb{C}$). Physicists work with the complex field. For simplicity, and because it is sufficient to establish our results, we shall work with the real field $\mathbb{R}$, although
everything goes with minor changes for the complex case. Let
(\textperiodcentered , \textperiodcentered )\ denote the scalar product in
Hilbert space $H$. We take $H$ of finite dimension $n$.

We shall be interested not so much in the Hilbert space as in operators, namely linear mappings $A:H\rightarrow H$. Such an operator $A$ is
Hermitian (or symmetric if we work over $\mathbb{R}$) if $(Ax,y)=(x,Ay)\ $for all $%
x,y\in H$. A Hermitian operator $A$ is non-negative if $(Ax,x)\geq 0\ $for
any $x\in H$\footnote{%
General Hermitian operators play the role of classical random variables. For any Hermitian operator $A$ and opinion state $O$, one can define
the `expected value' of $A$ as ${\rm Tr}(AO)$.}.\par \medskip

\textit{Opinion state}\newline
Each citizen $j\in \Omega$ is characterized by her opinion state and her thinking frame or perspective.  With the help of the trace one can introduce the notion of state of a quantum system. The trace ${\rm Tr}$ of a matrix can be defined as the sum of its
diagonal elements. It is known that the trace does not depend on the choice
of basis. Let $o$ (for opinion) be an element of $H$ with length 1 (that is $%
(o,o)=1$). Let $P_{o}$ be the orthogonal projector\footnote{%
A projector is an Hermitian operator such that $P^{2}=P$.} on $o$, that is $%
P_{o}(x)=(x,o)o$ for any $x\in H,\ {\rm Tr}(P_{o})=(o,o)=1$. So $P_{o}$ defines 
a state which we denote as operator $O(=P_{o})$. Such states corresponding to one-dimensional projectors are called pure states\footnote{%
As opposed to more general mixed states, not considered in this work. Pure state are complete information states, but because of intrinsic indeterminacy, the values along alternative (incompatible) perspectives can never be sorted out simultaneously, they remain stochastic.}. The non-negativity of the operator $O$ is analogous to the non-negativity of a probability measure, and trace 1 analogous to the sum of probabilities which equals 1. This means that an opinion state is formally identical to a (subjective) belief state \cite{dan18b}.\\

\textit{Perspectives}

The formal account of thinking frames is a key building block of our theory. It is intimately linked with our cognitive assumptions: 
\begin{assumption} ~
     \begin{enumerate}
    \item[i] Citizens cannot address reality immediately. They need to build a \textit{representation} of the voting issue using a thinking frame;
    \item[ii] Citizens cannot resort to a ``super frame" that would aggregates all relevant aspects.
\end{enumerate}
\label{assum_cog}
\end{assumption}

Assumption 2 implies that, for the citizens, the voting issue admits several
equally valid but Bohr-complementary thinking frames, the \textit{perspectives}. Formally, the experts and citizens define a set of perspectives (or thinking frames, we use the
terms interchangeably). A perspective ${\cal P}=\left( P_{1},...,P_{n}\right)$ is a collection of pairwise orthogonal one-dimensional projectors (${\rm Tr}(P_i P_j)=\delta_{i,j}$ the Kronecker delta) with the property $\sum_{i=1}^n P_{i}=E$, where $E$ is
the identity operator on $H$. Perspective ${\cal P}$ is incompatible with perspective ${\cal Q}=(Q_1, \dots, Q_n)$ if $P_i Q_j \neq Q_j P_i$ for some of the $i,j$. Correlations among perspectives are captured by the scalar products ${\rm Tr}(P_i Q_j)$. Incompatibility of ${\cal Q}, {\cal P}$ is then equivalent to ${\rm Tr}(P_i Q_j) \notin \{0,1\}$ for some of the $i,j$.\\

\textit{Probing a perspective }

Probing a perspective is analogous to performing a measurement in QM. It is formalized as an operation whereby one applies
some perspective ${\cal P}$ to an opinion state $O$. In deliberation terms, it
corresponds to questioning oneself in the terms of perspective ${\cal P}$. For instance, in the environmental perspective: ``Do I think that ICB is an efficient way to reduced Green House Gas emissions or not?" We focus exclusively on \textit{%
complete measurements}, namely on probing operations that fully resolve
uncertainty with respect to the ${\cal P}$ perspective. The outcome of probing a
perspective is one of the labels $i\in\{1, \dots, n\}$, here corresponding e.g. to the opinion: ``I believe ICB is efficient to reduce GHG".

The probability $p_i$ to obtain outcome $i$ in opinion state $O$ is equal to ${\rm Tr}(P_i O)$, and if $i$ is obtained, the revised opinion state is given by $P_i$. Hence, if the probing of the same perspective is repeated, as ${\rm Tr}(P_i P_j)=\delta_{i,j}$, the same outcome $i$ is obtained again and the opinion state does not change. An important feature that we emphasize is that the expected revised opinion $O_{ex}=\sum_{i=1^n} p_i P_i = \sum_{i=1^n}P_{i}OP_{i}$ is generally different from the initial opinion state $O$.
That is, although the opinion state has the structure of a probability
distribution, the revised opinion in the quantum formalism is \textit{not%
} subject to Bayesian plausibility, as noted in \cite{dan18b}. This feature
plays an important role in the analysis.\medskip

For the sake of illustration, assume a citizen $j \in \Omega$ is characterized by her own thinking frame ${\cal P}$ and opinion state $%
O=P_{i}$ for some $i \in \{1, \dots, n\}$. Probing an argument in an \textit{alternative} (i.e.,
incompatible) perspective ${\cal Q}=(Q_1, \dots, Q_n)$ is called \textit{%
challenging one's opinion}. For instance, the individual is confronted with the
liberty perspective and reflects over whether she thinks that ICB is
contrary to fundamental individual liberties ($Q_{1}$), or implies acceptable
limitations ($Q_{2}$), or has no implication for individual liberty ($Q_{3}$)
Challenging one's opinion results in a new state $Q_j$, with probability ${\rm Tr}(P_i Q_j)$.\\

\textit{Updating one's opinion}

\begin{figure}
    \centering
    \includegraphics[width=1\linewidth]{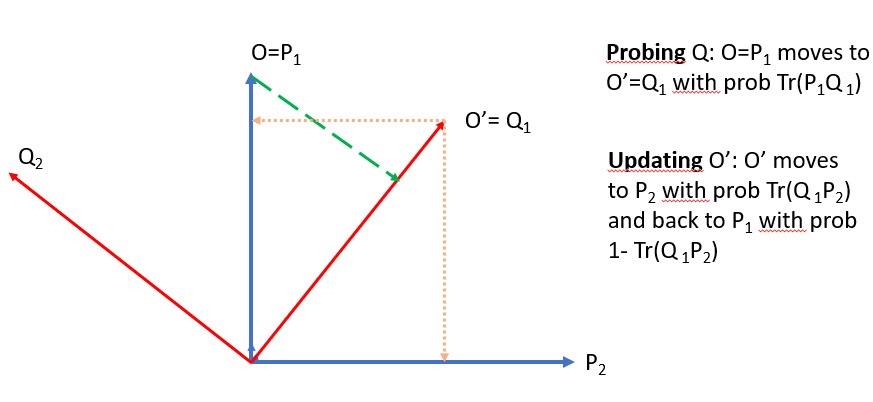}
    \caption{Probing perspectives}
    \label{fig_schema}
\end{figure}

In order to vote, the citizen has to come back to her own thinking frame (see more details below in Section \ref{sec_utility}), which corresponds to probing her own perspective ${\cal P}$. We call this process \textit{updating} one's opinion. A central feature of the model is that if the citizen started from opinion state $P_i$, corresponding e.g. to the opinion that ICB are environmentally beneficial, probing perspective ${\cal Q}$ and then updating her opinion might result in a different opinion state $P_j$ with $j \neq i$. Indeed, opinion state $P_j$ is reached with the overall probability ${\rm prob}(P_i \to P_j)= \sum_{k=1}^n {\rm Tr}(P_i Q_k) {\rm Tr}(Q_k P_j)$, which is nonzero in the general case for $j \neq i$. In deliberation terms, the citizen might now have the opinion that ICB are environmentally useless. This is the crucial property that generates a potential for opinions to evolve without additional information.
This is illustrated in Fig.~\ref{fig_schema} in the two-dimensional case. With initial opinion state given by $P_{1}$, the citizen probes the liberty perspective ${\cal Q}$. With probability ${\rm Tr}\left( P_{1}Q_{1}\right)$ she finds that in that perspective her opinion is $Q_{1}$ (and with the complementary probability it is $Q_{2}$). She then update her opinion and with probability ${\rm Tr}\left(
Q_{1}P_{2}\right)$ she does not recover her initial opinion: instead she now holds opinion $P_{2}$.

\subsection{Utility and voting}
\label{sec_utility}
Voting is a binary, Yes or No, choice. Citizens are endowed with preferences that allow them evaluating the (expected) utility value of the two options given their
individual opinion state.

\begin{assumption}
A Citizen can only attribute utility value to voting options in one of the eigenstates of her own frame i.e. in some state $O=P_i$ with ${\cal P}=(P_1, \dots, P_n)$ the citizen's \textit{own} perspective. 
\label{assum_util}
\end{assumption}

Assumption \ref{assum_util} captures a central feature of thinking in frames. A citizen is
only capable of evaluating their utility in terms of her own perspective. Citizens can explore other perspectives and adopt any
possible opinion. However, an opinion state can guide voting \textit{only}
when formulated in terms of the own perspective. The frame is an essential
part of the citizen's identity, it is the language in which they can
formulate the value of options and make decisions.

Formally, consider a citizen with perspective $\mathcal{P}=(P_{1},...,P_{n})$
Her utility function is represented by a tuple $\left\{
u_{1}^{Y},..,u_{n}^{Y},u_{1}^{N},...,u_{n}^{N}\right\} $ which associates a real number to option Yes respectively No to each possible opinion state. Generally, for any arbitrary opinion state $O$, one can formulate the expected utility for the two voting options as a ``quantum lottery" \cite{dan18c}, with the probability for each of the states $P_{i}$ given by ${\rm Tr}(OP_{i})$. The general formula for the expected utility is then 
\begin{equation}
\mathrm{EU}\left( Y;O\right) =\sum_{i=1}^n \mathrm{Tr}\left( OP_{i}\right)
u_{i}^{Y}  \label{EUdef}
\end{equation}
and similarly $\mathrm{EU}\left( N;O\right) =\sum_{i=1}^n \mathrm{Tr}\left(OP_{i}\right) u_{i}^{N}$. When $\mathrm{EU}\left( Y;O\right) \geq \mathrm{EU} (N;O)$, the citizen prefers the Yes option. Note that in order for the expression in Eq.~\eqref{EUdef} to guide voting behavior, the citizens must be able to compute ${\rm Tr}\left( O_{{}}^{{}}P_{i}^{{}}\right) $ for any $O$. This is very demanding. In particular, it would require that citizens have knowledge about the correlations between all alternative perspectives with their own. We do not make that assumption. Instead, whenever in $O \notin {\cal P}$, the citizen has to update their opinion state by probing their own thinking frame before voting.

\begin{assumption} 
 A citizen is endowed with a thinking frame that allows separating between voting options.
 \label{assum_diff}
\end{assumption} Assumption \ref{assum_diff} excludes citizens whose voting decision is fixed and therefore cannot be affected by deliberation. For any citizen $j$ with perspective ${\cal P}^j=(P_1^j, \dots, P_n^j)$, some $P_i$ give rise to a Yes vote, while some $P_j$ give rise to a No vote. The voting behavior is most simple. We know by Assumption \ref{assum_util} that citizen $j$ must be in one of her perspective's eigenstates, say $P_i^{j}$. Then if $u_i^Y \geq u_i^N$, citizen $j$ casts a Yes vote; otherwise she casts a No vote, i.e., voting is
sincere (non-strategic)\footnote {Notice that the utility function is in general different for each citizen $j\in \Omega$.}. The winning option is the one that has obtained the
largest number of votes. In case of tie at the end of the deliberation, a random draw determines the decision. 

\subsection{Deliberation}

Deliberation is modeled in terms of a multiple-round mediated communication
game. The facilitator is the sole true player in this game. By Assumption \ref{assum_0},
the citizens follow the facilitator's recommendations. In each round, the facilitator chooses a perspective ${\cal P}$ that will be exposed by a citizen or an expert, and a selection of citizens (defined as
being \textit{active}) whom he invites to probe the corresponding perspective 
${\cal P}$. All other citizens remain ``passive": they only listen and thus do not change opinion. When hearing an
argument, an active citizen explores ${\cal P}$  by
thinking in its terms, so as to determine which argument
she agrees with.\footnote{
This is more demanding than simply deciding to agree or reject the presented argument. Indeed if the argument is rejected, the revised state would not be an eigenstate of the perspective under exploration but a superposition of states.}
After challenging her opinion by probing some ${\cal P}$, the citizen updates her
opinion by reassessing her position in her own perspective where she can
evaluate the utility value of the voting options, as described above.\\

\textit{The facilitator strategy}

In each round the facilitator makes two choices: a perspective ${\cal P}$ (chosen among the citizen's and expert's perspectives) and $\omega \subset $ $\Omega $ ($\Omega=\{1, \dots, N\}$ is the set of all citizens), the set of citizens who are invited to be active. The facilitator's objective is to maximize the support
for the most supported voting option. Let $y^{t}\left( n^{t}\right) $\ be the number of Yes(No)
supporters. Deliberation is a finite-period game aimed at maximizing the
probability for consensus among citizens before period $T,T\geq 1.$
 We operationalize this with a "score function" $s^{t}\left(
.\right) =\max \left\{ y^{t},n^{t}\right\} ,\ s^{t}\left( .\right) $\ is
simply the score of the most supported option in period $t$. \par 

Next, in a multi-period game, a strategy describes what to do in each
period $t\leq T-1\ $after each possible history. A crucial remark is
that the vector of opinion states $\mathbf{O}^{t}=(O_1^t, \dots O_N^t)$ captures all relevant information about
history at time $t$. This follows from the fact that the probability for citizen's $j$ opinion change when probing perspective ${\cal P}=(P_1, \dots P_n)$ is captured by ${\rm Tr}(O_j P_{i})$ i.e., only depends on the current state of opinion -- not on previous history. Finally, we shall restrict our attention to strategies that maximize the expected score \textit{in each period}. Notice that this is not fully without loss of generality.\footnote{In \cite{lambertF24} we show that under some circumstances, optimizing over several rounds suggests different optimal strategies.} Hence, a strategy is defined as a function from a vector of opinions $\mathbf{O}^{t}$ to a pair $( \cal {P},\omega )$ including a perspective $\cal{P}$ and a set of active citizens $\omega :$ $\mathbf{O}^{t}\rightarrow \mathcal{P}\times \left( 2^{N}-1\right)$, where $2^N-1$ counts the number of non-empty sets of $\Omega=\{1, \dots, N\}$. The facilitator objective is to maximize the expected support: $\left\{ \max \left\{ Ey^{t},En^{t}\right\} \right\}$ in each period.

We make the following informational assumption:

\begin{assumption}
The facilitator has knowledge about all perspectives and about the correlations between them.
\end{assumption}

The facilitator has an informational advantage compared to citizens in terms
of knowing the full structure of the opinion state space. He is aware of all
possible perspectives and how they correlate to each other. This is the
critical resource that allows him to optimize the deliberation process.
We also sometimes consider that he has access to a pool of experts who can present arguments from any possible perspective ${\cal P}^{e}$. Experts are simply "tools" that the
facilitator can call upon whenever he wants. However, since the spirit
of deliberation is also to give voice to citizens, we shall give
particular attention to what can be achieve without appealing to experts.

\section{Analysis}

We start the analysis with deliberation between
two citizens. This allows deriving our main results which we later extend to deliberation involving larger groups of citizens.

\subsection{Deliberation with 2 citizens}
\label{sec_2citizens}
We have 2 citizens Alice and Bob who face a binary collective decision about e.g.,
the introduction of an Individual Carbon Budget scheme. There exist two
relevant aspects: environmental efficacy and individual liberty. These 2
aspects cannot be considered simultaneously by our citizens, they are
assumed to be incompatible in the mind.

The perspectives are $n-$dimensional: the larger $n,$ the finer the
characterization of the opinions. In terms of our example, the
environmental perspective can have several possible values (each associated
with its own eigenstate) e.g., ICB is the best solution for reducing GHG
(greenhouse gas), ICB is one among the best solutions to reduce GHG, ICB
is a good solution etc., until ICB is worthless to reduce GHG. For the ease of presentation we shall focus on the case $n=2$ which allows establishing some central results (for more general results see our article \cite{lambertF24}). 

The citizens are endowed with
\begin{description}
\item[i.] An opinion state $O^{A(B)}$;
\item[ii.] An own perspective. Denote ${\cal A}=(A_1, A_2)$ Alice's perspective and ${\cal B}=(B_1, B_2)$ Bob's perspective;
\item[iii.] A set of utility values associated with the own perspective's
eigenstates: $\left\{
u_{A_1}^{Y},u_{A_2}^{Y},u_{A_1}^{N},u_{A_{2}}^{N}%
\right\} $, $u_{A_i}^{Y}$ is the utility value  for Alice
corresponding to the Yes vote in opinion state $A_{i}$, and $%
u_{A_i}^{N}$ corresponding to the No vote. We similarly define $%
u_{B_j}^{Y}, u_{B_j}^{N},$ $j=1,2$ for Bob.\medskip
\end{description}

\textit{Deliberation protocol.--} Before Bob and Alice start deliberating,
they update their own opinion, namely they probe their own perspective\footnote{This is most natural since they are expected to be invited to expose their opinion}. The
initial opinion states are therefore always eigenstates of the own
perspective: some $A_{i}$ for Alice and some $B_{j}$ for Bob. We consider a process where the facilitator does not
appeal to experts in the first 2 rounds (we call it the "voice phase").

\emph{Timing}

$t=0$ The facilitator asks for initial voting intentions. If Alice and
Bob agree, the procedure requires no deliberation. If they
disagree the first round starts:

\textit{Round 1}

$t=1$\ A random draw determines who will present his/her argument first, say
Alice;

$t=2$\ Alice's argues for Bob in her terms (frame) and Bob\ responds by
probing Alice's perspective;

$t=3\ $Bob updates his opinion by probing his own perspective. The facilitator asks agin for voting intention and if disagreement remains, the second round
starts:

\textit{Round 2}

$t=4$ Bob exposes his opinion in his own frame and Alice responds by probing
Bob's perspective;

$t=5\ $ Alice updates her opinion by probing her own perspective.

$t=6$ Out of the resulting opinion states, if they
still do not agree, the process 1-3 is repeated - possibly appealing to experts.

$t=8$\ if disagreement persists at $t=T$, the decision is determined
by a random device.

Alices's and Bob's respective opinion states are independent. The global
system is characterized by the following 4 possible (combined)
opinion states {${\bf O} \in \left\{ (A_{1},B_{1}), (A_{1},B_{2}), (A_{2},B_{1}),  (A_{2},B_{2}) \right\}$}. By
Assumption \ref{assum_diff} and since Alice and Bob only have two possible opinions, we
may, without loss of generality, define $A_{1}$ and $B_{1}$ as the opinion
states leading to a Yes vote; and $A_{2}$ and $B_{2}$ as the opinion states
leading to a No vote. Agreement on voting is then entirely defined by the
opinion states without the need to refer to utility values.  In both consensual cases (namely $(A_1,B_1)$ and $(A_2,B_2)$) the
facilitator's score is maximal and equal to 2. When the vector of opinion states $\mathbf{O} \notin \left\{ \left( A_{1},B_{1}\right) ,\left( A_{2},B_{2}\right)
\right\}$, the facilitator needs one of the citizens
to change her opinion. Maximizing the expected score is equivalent to
maximizing the probability that this happens. In the two-dimensional case,
the probability for Alice or Bob to change opinion when probing the other's
perspective is entirely governed by a single parameter: $x=\mathrm{Tr}%
(A_{1}B_{1})$. Indeed, as $A_{1}+A_{2}=E=B_{1}+B_{2}$, with $E$ the identity
operator, we have $\mathrm{Tr}(A_{2}B_{1})=\mathrm{Tr}(A_{1}B_{2})=1-x$, and 
$\mathrm{Tr}(A_{2}B_{2})=x$.

We consider deliberation starting from initial disagreement, say Alice and
Bob have respective opinion states $A_{1}$ and $B_{2}$. From the facilitator's point
of view, the two consensus states $(A_1,B_1)$ and $(A_2, B_2)$ are fully symmetric. Let the initial
random draw give Alice the initiative. The facilitator invites her to
present her argument. Bob is then invited by the facilitator to probe
Alice's perspective, and his opinion state changes to $A_{1}$ with
probability $\mathrm{Tr}(A_{1}B_{2})=(1-x)$, and to $A_{2}$ with probability $%
\mathrm{Tr}(A_{2}B_{2})=x$. Bob then updates his opinion by probing his
own perspective. The probability for reaching consensus is the probability
that Bob's final state is now $B_{1}$, instead of his initial state $B_{2}$.
This state is reached with probability: 
\begin{eqnarray}
\mathrm{prob}(B_{2}\rightarrow B_{1}) &=&\mathrm{Tr}(B_{2}A_{1})\mathrm{Tr}%
(A_{1}B_{1})+\mathrm{Tr}(B_{2}A_{2})\mathrm{Tr}(A_{2}B_{1})  
\\
&=&2x(1-x) ~.\label{Con2X2}
\end{eqnarray}%
Since only Bob's opinion has been challenged, the other consensual state
(namely $(A_{2},B_2)$) cannot have emerged.\newline

\textit{Maximally-uncorrelated perspectives.--} Generally, 
perspectives ${\cal A}$ and ${\cal B}$ are said to be maximally uncorrelated when $\mathrm{Tr}%
(B_{i}A_{j})=1/n$ for all $i,j$ (with $n=2$ in the present case). In this case, probing the $\mathcal{A}$ perspective
gives Bob equal chance to move into any of the states $A_{i}$ (here, Bob
reaches $A_{1}$ or $A_{2}$ with probability $1/2$). Updating then his
opinion by probing the $\mathcal{B}$ perspective, Bob reaches any of the
opinion states $B_{j}$ with equal probability. Effectively, Bob's initial
opinion state has been completely randomized by the (intermediate) probing of
Alice's completely uncorrelated perspective $\mathcal{A}$. In such a case,
starting from initial disagreement, the chance for reaching consensus after
one round is equal to $1/2$. \ 

The second round following disagreement
proceeds similarly, generating a probability for consensus in state $%
A_{2}B_{2}$ with the same probability $2x(1-x)$. Hence after the two rounds
the probability for reaching consensus is $2x\left( 1-x\right) +\left(
1-2x\left( 1-x\right) \right) 2x\left( 1-x\right) =4x\left( 1-x\right) \left[
1-x\left( 1-x\right) \right] $.

We have the following

\begin{proposition}
\label{prop1}
Starting from disagreement on vote between two citizens:

i. fact-free deliberation between citizen who share the same formal perspective (the two perspectives are fully correlated) has no impact on their opinions;

ii. with distinct perspectives consensus is reached with strictly positive
probability after a first round;

iii The probability for consensus is largest when the perspectives are
maximally uncorrelated, it reaches $3/4$ after two rounds.
\end{proposition}

\begin{proof}
i. When the two perspectives are fully correlated, we have $\mathrm{Tr}%
(A_{i}B_{j})\in \left\{ 0,1\right\} $: the opinion state are either equal ($%
A_{1}=B_{1}$ and $A_{2}=B_{2}$) or orthogonal ($A_{1}=B_{2}$ and $%
A_{2}=B_{1} $): Bob's and Alice's perspectives are formally
indistinguishable. In this case, no transition $B_{1}\rightarrow
A_{i}\rightarrow B_{2}$ or $B_{2}\rightarrow A_{i}\rightarrow B_{1}$ can
ever occur by probing the intermediate $\mathcal{A}$ perspective: letting
Bob probe Alice's perspective has no impact whatsoever on Bob's opinion
state ($2x\left( 1-x\right) =0$). Initial disagreement cannot be overcome
through deliberation.

ii. First note that from the point of view of the facilitator, the two
consensus states are fully symmetric $\mathrm{prob}(B_{2}\rightarrow B_{1})=%
\mathrm{prob}(A_{1}\rightarrow A_{2})=2x\left( 1-x)\right) $ so a random
draw is optimal for the facilitator. The result in 1.ii. follows from Eq.~%
\eqref{Con2X2}, which shows that the probability to reach consensus is
strictly positive whenever $0<x<1$, namely when perspectives $\mathcal{A}$
and $\mathcal{B}$ are distinct.

iii. The probability for consensus in the first round
is\ maximal at $x=1/2$ where $\ 
\frac{\partial }{\partial x}[2\left( 1-x\right) x]=0,x\in \left[ 0,1\right]
.\ $The total probability for success after the second round, conditional on
failure in the first round, is $4x\left( 1-x\right) \left[ 1-2x\left(
1-x\right) \right] $ which reaches its maximum for uncorrelated perspective
as well i.e., $x=1/2$. For two rounds the maximum chance for consensus is $%
3/4$. 
\end{proof}
\medskip \ 

Proposition \ref{prop1}.i. is quite remarkable because it shows that starting from
disagreement, sharing the same
thinking frame is an obstacle to achieving consensus. Indeed, within a common thinking frame, citizens can only
update their opinion in response to new information (by Bayesian updating)
which we preclude in this paper. However, when citizens are endowed with
distinct perspectives new opportunities for opinion to evolve arise. By
actively exploring a perspective incompatible with ones' own, intrinsic
contextuality reveals its transformative power. Exploring an alternative
perspective changes the opinion state because the possible outcomes of that
operation do no exist in the own perspective. The opinion state is forced
into a new state. This result about the value of diversity is truly novel
and a main contribution of this paper. It is important at this point to recall \ref{assum_0}, the transformative value of
deliberation demands a true mental experience i.e., sincerely "putting
oneself in someone else shoes" - the probing operation. In the context of the example, this means for Alice to think in terms of individual liberties, their significance for society, general risks linked to  liberties etc.... So the idea is to sincerely recognize that Bob has a point in bringing up this aspect even if Alice may disagree with his opinion. 

Result 1.ii quantifies how the diversity of perspectives allows opinions to
evolve toward consensus. The weaker the correlation between perspectives$\
x\rightarrow 1/2$, the larger the impact of the probing operation. Starting from
disagreement $(A_{1},B_{2})$, the probability that Alice changes her opinion
from $A_{1}$ to $A_{2}\ $is given by ${\rm Tr}\left( A_{1}B_{1}\right) {\rm Tr}\left(
B_{1}A_{2}\right) +{\rm Tr}\left( A_{1}B_{2}\right) {\rm Tr}\left( B_{2}A_{2}\right)
=2x\left( 1-x\right) $ which tends to zero as $x$ tends to 1(or 0) and it is
maximized at $x=1/2$. The intuition is that the more closely related the perspectives the more
likely that probing Bob's frame takes Alice to an opinion state (in ${\cal B}$) close to her initial state (when $A_{1}$ is close to $B_{2},{\rm Tr}( A_{1}B_{2})$ is large).  When probing her own perspective anew she is most likely
confirmed in her initial opinion. Similarly for Bob, so disagreement is more
likely to persist. Nevertheless with some positive probability one
of the two citizens will end up having changed his or her mind, which implies
consensus on voting. Interestingly, the result that uncorrelated
perspectives give the best chance for deliberation to achieve consensus,
reminds of a result in quantum persuasion \cite{dan18b}, where the authors show
that distraction understood as bringing attention to a perspective
uncorrelated to the targeted state is part of an efficient strategy to persuade Receiver.

Result iii. says, without surprise, that starting from dissensus, additional
rounds following failure to reach agreement increases the probability for
consensus. While a single round can already achieve consensus with
probability $2\left( 1-x\right) x,$ with two rounds and maximally uncorrelated
perspective case ($x=1/2)$, we reach consensus in 75\% of the case. Of course
we do not expect citizens to repeat the same argument from round to round: if
citizens are short of arguments, it can be relevant to call for experts with suitable alternative perspectives. \medskip\ \ 

\begin{corollary}
The first moving citizen has larger chance to see consensus on her initial
voting preferences than the one who moves second.
\end{corollary}

\begin{proof}
As earlier noted that the chance of reaching consensus is the same whoever is selected first: $%
{\rm prob}\left[(A_{1},B_{2}) \to (A_{1},B_{1})\right] =2x\left( 1-x\right) ={\rm prob}\left[(A_{1},B_{2}) \to (A_{2},B_{2})\right]$. But since the second case is only possible in case
of failure in the first round, it has less chance to
be selected in the vote.
\end{proof}

The procedure gives more chance to the first selected citizen. The
introduction of a random draw restores the equality of chance between
citizens. \medskip

\begin{corollary}
When citizens's perspectives are correlated, relying on experts with
perspective uncorrelated to the current round's active citizen increases the
chance for reaching consensus in any given round.
\end{corollary}

This follows from result ii. When Alice and Bob have correlated perspectives,
the probability for reaching consensus when probing each other's perspective
is lower than 75\% after two rounds. The facilitator could choose instead
the following strategy. First, a random draw designates the active citizen. The procedure then proceeds as above except that only
experts are presenting arguments belonging to a perspective maximally uncorrelated
with the active citizen's.

Corollary 2 implies that there exists a tension between letting citizen
expose their argument and probe each other's perspectives, and the objective to maximize the score. This is not surprising given
Proposition \ref{prop1}.ii. 
In order to preserve the democratic character of deliberation, and to give
voice to citizens, a mixture of citizen's arguments and expert arguments can be
chosen, at the cost of some delay
in reaching consensus. For instance, deliberation could start with a voice phase where citizens present their own views, and if it fails proceed with an expert phase.

\subsection{Deliberation in a population of citizens}
\label{sec_population}

We next analyze deliberation in a population of voters akin to a citizen
assembly. We consider the simple but most relevant case where the population is divided into two groups each, with its own thinking frame
about the voting issue. The ICB example is a suitable one as it relates to
quite well-established ideologies: a left leaning social and environmental
ideology (${\cal L}=(L_1, L_2)$), and a right leaning conservative libertarian ideology (${\cal R}=(R_1,R_2)$). We assume that $L_1$ and $R_1$ give rise to a Yes vote, while $L_2$ and $R_2$ give rise to a No vote. Hence the two consensual state are $(L_{1},R_{1})$ and $(L_{2},R_{2})$. To ease comparison with the previous section we define ${\rm Tr}(L_{1}R_{1})={\rm Tr}(L_{2}R_{2})=x$. 

We first establish a simple Lemma:

\begin{lemma}
Starting from consensus, one round of deliberation between two citizens with
distinct perspectives leads to disagreement with positive probability.    
\end{lemma}

\begin{proof}
Consider starting from $(L_{1},R_{1})$ and letting the ${\cal R}$-citizen probes the ${\cal L}$
perspective, we are back in $(L_{1},R_{1})$ with probability $%
{\rm Tr}(R_{1}L_{1}){\rm Tr}\left( L_{1}R_{1}\right) +{\rm Tr}(R_{1}L_{2}){\rm Tr}\left(
L_{2}R_{1}\right) = x^{2}+(1-x)^{2}<1$, so consensus is lost with
positive probability for $x\notin\{0,1\}$ that is when the two perspective are distinct from each other. 
\end{proof}

Lemma 1 assesses that when two citizens agree on voting, it can be harmful for the existing consensus to let them explore alternative perspectives as it may trigger a change in opinion.  As we show below, this feature has implications for the optimal strategy
of the facilitator in the population context. We note that it is
consistent with much criticism appealing to cognitive biases that emphasizes
that interacting politically can have ``negative" impact on citizens' beliefs
and preferences.(\cite{janis82}, \cite{Elster05}). \medskip

Let the two groups be of size $l$, respectively $r$. We shall assume that in
each group the citizens have determined themselves in their own frame
among the 2 possible opinions, so we have two distributions $\left(
l_{1},l_{2}\right) ,\ l_{1}+l_{2}=l$ and similarly $\left(
r_{1},r_{2}\right) ,\ r_{1}+r_{2}=r$. A new feature compared to the previous
analysis is that there can be disagreement among people sharing the same
perspective, i.e., between the ${\cal L}({\cal R})$ people supporting the Yes vote, $%
l_{1}(r_{1})$, and those who support the No vote, $l_{2}(r_{2})$. 
 We assume, without loss of generality, that $l_{1}+r_{1}>l_{2}+r_{2} $, so that there is an initial  majority for the Yes vote. The objective of the facilitator is to maximize the score in the next round. The first step is to select the \textit{projected consensus state}, defined as the consensus state that the facilitator aims at maximizing support (score) for.  The next step is to select the citizens who will be invited to probe a perspective. By Lemma 1, we know that only citizens disagreeing with the projected consensus should be invited to be active. Otherwise, the facilitator risks losing some support for his projected consensus. In general, there are disagreeing citizens in both ${\cal L}$ and ${\cal R}$ perspective. The probability for one citizen of switching voting intentions when confronted to the alternative thinking frame is equal to $2x(1-x)$. The gains in terms of score from a deliberation round are therefore proportional to the size of the group (of disagreeing citizens) selected for probing. As we shall, counter-intuitively, see this may imply selecting the minority consensual state as the projected consensus. We have the following proposition: 

\begin{proposition}
\label{prop2}
Starting from disagreement within and/or between 2 groups, the optimal strategy for the facilitator in each round entails:\newline
i. Choose the projected consensus state as the one associated with largest expected score under optimal choice of target group and perspective;\newline
ii. Selective targeting: Select the largest group of citizens disagreeing with the projected consensus for probing and demand that remaining citizens from that perspective group refrain from probing (only listen);\newline
iii. c\ 
\end{proposition}

\begin{proof}
2.i. We know from Proposition 1 that the probability for opinion change is $2x(1-x)=\Delta$. The expected change is thus $\Delta l_{i}$ (or $\Delta r_{j}$) where $i,j=1,2$ depending on the group invited for probing. Consider a case where $(l_{1}+r_{1})>(l_{2}+r_{2})$, so the majority supports Yes. Then whenever $l_{1}+r_{1}+\Delta \max(l_{2}, r_2) < (l_{2}+r_{2})+\Delta \max(l_{1}, r_1)$, that is $(l_{2}+r_{2})-(l_{1}+r_{1})>\Delta[\max(l_{2},r_2)-\max(l_{1}, r_1)]$, the optimal projected state corresponds to the No (minority) vote. When the inequality goes the other way the standing majority is the optimal projected consensus state.  2.ii Given 2.i the largest group of citizens disagreeing with the projected consensus is identified. Because of Lemma 1, no other citizen from that perspective group should be invited to probe.  
2.iii Let $l_{1}=l_{2}=l/2$ and similarly $r_{1}=r_{2}=l/2$ with $\Delta=1/2$. Any of the two consensus states can be reached with the support of a population of size $l/2+r/2+l/4+r/4=(3/4)(l+r)$.  \medskip
\end{proof}

    The results in Proposition 2 invite multiple remarks. 
First, in each period the projected consensus state is not necessarily the standing majority. This is because we are dealing with a population of citizens, so the size of the group that could switch opinion matters. While the majoritarian consensual state holds an advantage, optimal deliberation can favor the minority consensus state. This is quite remarkable because the result applies already when the facilitator is ``myopic" i.e., maximizes the score at each step. This result is a nice property of the deliberation procedure, as it implies that it gives a fair chance to the minority\footnote{In our general paper we show that overturning the initial majority can also be optimal when the facilitator maximizes consensus probability over several rounds \cite{lambertF24}.}. Let us illustrate with a numerical example: \medskip 

\textbf{Example}
Consider the following initial situation with $\Delta=1/2$: $l_{1}=60,\hspace{0,2cm}  l_{2}=40$ and $r_{1}=20,\hspace{0,2cm}  r_{2}=35$ implying $l_{1}+r_{1}=80 > l_{2}+r_{2}=75$, so the standing majority is $l_{1}+r_{1}$. If the facilitator selects $(L_{2},R_{2})$ as the projected consensus state, after one round the expected score is $l_{2}+r_{2}+\Delta l_{1}=105$ which is larger than the score he can reach with the standing majority ($l_{1}+r_{1}+\Delta l_{2}=100$), so the facilitator should better lead deliberations that may eventually overturn the initial voting intentions. \medskip 

 A second important remark is that the optimal strategy actualizes the distinction between active and passive citizens. Because of Lemma 1, when both opinions are present in a perspective group, it is optimal for the facilitator to proceed selectively. Those from the targeted perspective group already agreeing with the projected consensus state should refrain from probing, as they could change their mind and reduce the score. This means that the optimal procedure implies to some extent the unequal treatment of citizens, which is unfortunate from a democratic ideal point of view. Note however that full publicity of debates can be preserved because with contextual opinions, only the operation of probing can induce change. Simply listening to an argument without making the effort of thinking in the terms of the alternative perspective has no effect on opinions.  

 Our analysis of the population case provides an interesting rationale (and guiding principles) for the practice of parallel working groups encountered in real life deliberations. We could say that it responds to the awareness that there exists risks that deliberation ``breeds confusion" in people's mind. Returning to the illustration, we could have a single round with two parallel working groups, one made up of  $l_{1}$ and the other of $r_{1}$. \footnote{ Note that $r_{2}$ citizen would be free to join the first group and $l_{2}$ the second working group. since they would be probing their own perspective that would not affect their opinion.}  Such a procedure will prevent unwanted opinion switch when citizens make probing operations without being invited to. Subsequent rounds could appeal to experts exposing arguments from maximally uncorrelated perspective to remaining disagreeing citizens. Obviously this speeds up the process toward consensus. 

We thus find that our results from Proposition 1 carry over to populations of citizens. Unexpectedly, we also find that the initial distribution of opinions does not fully determine the outcome of voting which is a nice property. Less attractive from a democratic point of view is that citizens are not treated equally. Given assumption \ref{assum_0}, we propose that the necessary discrimination can be implemented less controversially when the probing operations are carried out in well-composed working groups.

\section{Concluding remarks}
\label{sec_discussion}
In this paper we have developed a formal approach to deliberation based on
the behavioral premises and the mathematical formalism of quantum cognition. Deliberation is formulated in the context of complete information as a
structured communication process managed by a facilitator with the aim of
maximizing the probability for consensus in a binary collective choice problem. The process
includes a sequence of rounds in which some citizen (or expert) develop an
argument belonging to some perspective and other citizens are invited by the
facilitator to probe that perspective. Probing involves a true mental experience for the
citizens. They have to think in the terms of the probed perspective and decide
how they position themselves.

A first central result is that the incompatibility of perspectives, that is
the diversity of viewpoints among citizens, is what permits
opinions to evolve. Our second central result is that the correlation
between perspectives is the key property that determines the pace of
evolution towards consensus. In the two-citizens case, starting from disagreement, the highest
probability for consensus is achieved when the citizens' perspectives are maximally uncorrelated. The results generalize to populations of citizens where the facilitator's
strategy involves some inequality in treatment: only a selected subset of citizens is invited to perform the probing operation, while the others simply listen. The population case reveals that optimal deliberation may overturn initial majority, which gives some chance to the initial minority option. 

We thus find that the quantum cognition approach delivers a transformative character of deliberation going beyond Bayesian updating. In our model, people go through real
(mental) experiences (probing) which transform their opinion. While in the model they always are in a
situation of complete information, they learn how the issue at stake
can be looked at from equally valid alternative perspectives and how these
relate to their own. This resonates with the normative imperatives often put
forward by deliberative democrats that deliberation requires engagement and fosters the respect for others as a practical school of democracy. In our context this is
captured by the willingness and actual experience of ``putting oneself in someone else's shoes". Our
approach also characterizes the determinant role of the facilitator who
does play a central role in most actual experiments with citizen assemblies. Not surprisingly, the optimal strategy of the facilitator features a tension between equity and efficiency translated into the recommendation for some citizens to be active while other should rather remain passive. Interestingly, this result can provide guiding principles for setting up working groups in parallel sessions. 
Finally and importantly, our analysis reveals a value of deliberation in terms of improving consensus not based on improved information. 
 \medskip

\pagebreak

\end{document}